\documentclass[preprint,1p,11pt]{TR_Templ}
\usepackage{flushend}
\usepackage{pgfplots,subcaption}
\newcommand{\meas}[2]{y_{#1}^{#2}}

\newcommand{\st}[1]{x_{#1}}

\newcommand{\xl}[2]{l_{#1,#2}}

\newcommand{\hnoise}[3]{h_{#1,#2}^{#3}}
\newcommand{\ymeas}[3]{y_{#1,#2}^{#3}}

\pgfplotsset{compat=newest}
\pgfplotsset{every tick label/.append style={font=\tiny}}

\usetikzlibrary{pgfplots.groupplots}
\usepackage{amsfonts}
\usepackage{amssymb}
\usepackage{mathrsfs}
\usepackage{amsmath}
\usepackage{amssymb}
\usepackage{mathrsfs}
\usepackage{dsfont}
\usepackage{url}

\usepackage{xargs}
\usepackage{color}

\definecolor{tcolor}{RGB}{0,60,104}
\usepackage{graphicx}

\newcommand*\vect[1]{\begin{bmatrix}#1\end{bmatrix}}
\newcommand*\tvect[1]{\big[#1\big]^T}

\newcommand{\mat}[1]{{\ensuremath{{\mathbf{#1}}}}}

\newcommand{\IR}{\NewR}
\newcommand*{\Exp}[1]{\E\hspace{-0pt}[#1]}
\newcommand*{\Cov}[1]{\Co\hspace{-0pt}[#1]}

\DeclareMathOperator{\Tr}{Tr}
\DeclareMathOperator{\E}{E}
\DeclareMathOperator{\Co}{Cov}

\newcommand{\NewR}{\ensuremath{\mathds{R}}}
\def\Eq#1{\eqref{#1}}

\def\Eq#1{\eqref{#1}}

\newcommand\Gauss[3]{{\cal N}(#1; #2,#3)}

\DeclareMathOperator{\diag}{diag}

\newcommand{\Fig}[1]{Fig.~\ref{#1}}

\begin{document}
\begin{frontmatter}
\title{Second-Order Extended Kalman Filter for\\ Extended Object and Group Tracking}

\author[data_fusion_goe]{Shishan~Yang}
\ead{shishan.yang@cs.uni-goettingen.de}

\author[data_fusion_goe]{Marcus~Baum}
\ead{marcus.baum@cs.uni-goettingen.de}

\address[data_fusion_goe]{Institute of Computer Science\\
University of G\"ottingen, Germany\\}

\begin{abstract}
In this paper, we propose a novel method for  estimating an elliptic shape approximation of a moving extended object that gives rise to multiple scattered measurements per frame.
For this purpose, we parameterize the elliptic shape with its  orientation  and the lengths of the semi-axes. We relate an individual  measurement with the ellipse parameters by means of a multiplicative noise model and derive a second-order extended Kalman filter for a closed-form recursive measurement update. The benefits of the new method are discussed by means of  Monte Carlo simulations for  both static and dynamic scenarios.
\end{abstract}

\end{frontmatter}

\section{Introduction}

Extended object tracking is becoming increasingly important in many application areas such as autonomous driving \cite{Granstroem2014a} and maritime surveillance \cite{Granstroem2014}.
An extended object is characterized by a varying number of noisy measurements from different spatially distributed sources on the object.
In contrast to point target tracking, the objective is to estimate both the location and  shape of the target object.
Typically, only  few measurements are available per frame so that it becomes necessary to systematically fuse measurements from different frames under incorporation of the temporal evolution of the object.

Many different extended object tracking methods with different properties and application areas have been developed in the past years. For a recent overview of extended object tracking and its applications, we refer to \cite{Granstroem2017}.
A  main challenge in extended object tracking is that joint tracking and shape estimation is a high-dimensional problem with severe nonlinearities, which requires sophisticated and problem-specific nonlinear estimation techniques.

One of the first approaches is the random matrix approach \cite{Feldmann2008,Feldmann2009,Feldmann2010,Feldmann2008a,Feldmann2012,Granstrom2010,Granstrom2012,Koch2008,Koch2005} that  models the spatial extent with a Gaussian distribution whose covariance matrix is recursively estimated. For this purpose, the uncertainty of the covariance matrix is represented with an inverse Wishart density.

The random hypersurface  (RH) model \cite{J_Baum2012,J_TAES_Baum,J_TAES_Baum_RHM,Fusion12_Faion-CylinderTracking,Fusion13_Faion,MFI15_Faion,Faion2014,Faion2015,Faion2015a}
 reduces the extended object tracking problem to a curve fitting problem by means of scaling the shape contours. This idea can be used for 
basic geometric shapes such as ellipses but also for general star-convex shapes and three-dimensional objects. In the RH approach the shape parameters are estimated using Gaussian estimators such as the Unscented Kalman Filter (UKF) \cite{Julier_UnscentedFiltering}.  Of course, in general, the increased flexibility comes at the cost of more complex algorithms. Monte Carlo methods for extended object and group tracking problems are described in \cite{Mihaylova2014,Petrov2012a,Petrov2011,Petrov2012}.

The  objective of this paper is to develop a \emph{Gaussian} state estimator, i.e., nonlinear Kalman filter, for the measurement model of the \emph{random matrix approach} \cite{Feldmann2012}.  

First, we define a  suitable parameterization of an arbitrary-oriented ellipse using the orientation and the length of the semi-axes. 
Second, we form a (polynomial) measurement function that relates a measurement to the state vector (including kinematic and shape parameters).
For this purpose, we follow the idea of our previous work \cite{Fusion12_Baum}, where a multiplicative noise is used to model the spatial extent of an extended object.
In order to perform a  closed-form measurement update based on the derived measurement equation, we derive a second-order extended Kalman filter (SOEKF) \cite{Roth2011}.

In contrast to the random matrix approach, the proposed method maintains the mean and joint covariance of the kinematic parameters, orientation, and lengths of the semi-axes. 
Hence, from a modeling point of view, the process model for the shape can directly work with the individual shape parameters.
Due to the standard Gaussian representation of the state vector our approach is easy to embed into multi-extended object tracking algorithms.

\begin{figure}
\centering
\includegraphics[scale=.85]{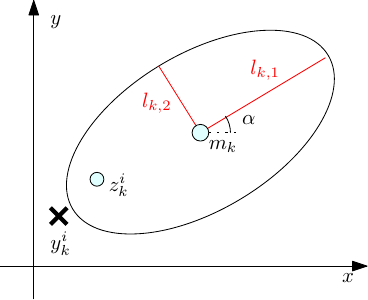}
\caption{Measurement model and ellipse parameterization. The $i$-th measurement source at time $k$ is $z_k^{i}$. Corresponding measurement $y_k^{i}$ is $z_k^{i}$ added with measurement noise. The object shape is modeled using state vector $x_k = [m_k^{T}, \alpha_k, l_k^{T}]^{T}$, where $m_k$ is the center, $\alpha_k$ indicates the object orientation, and $l_k=[l_{k,1},l_{k,2}]^{T}$ describes the size of the extended object.}
\label{fig:ellipse_par}
\end{figure}

\section{Extended Object Model}
This section introduces the state vector, measurement model, and process model used in this work for tracking a single extended object.
\subsection{State Vector and Shape Parametrization}
 The state vector
 \begin{equation}
 \st{k}=\tvect{m_k^T, p_k^T, r_k^T}  
 \end{equation}
 consists of both kinematic parameters, i.e., center  $m_k \in \IR^2$, possibly further quantities $ r_k \in  \IR^{n_r}$ (e.g., velocity),  and shape parameters  $p_k \in \IR^{3}$.
   We propose a parameterization of an ellipse according to 
   \begin{equation} \label{eqn:param}
p_k = \tvect{\alpha_k, \xl{k}{1}, \xl{k}{2} }\in \IR^3    
   \end{equation}
 where 
 \begin{itemize}
 \item   $\alpha_k \in [0, \tfrac{\pi}{2}]$ specifies the orientation at time step $k$, and 
\item  $\xl{k}{1} \in \IR^+$ and $\xl{k}{2} \in \IR^+$  specify the lengths of the semi-axis.
\end{itemize}
Note that this is an obvious and intuitive parameterization of an ellipse. For example, it has also been used in \cite{Granstroem2011} for a different measurement model.
Besides ellipses, this parameterization could be employed for other perpendicular axis symmetric shapes, e.g., rectangles.

\subsection{Measurement Model}
 At each time $k$, the extended object gives rise to $n_k$ (independent) two-dimensional measurements 
\begin{equation}
\mathcal{Y}_k = \{ \meas{k}{i} \}_{i=1}^{n_k} \enspace, 
\end{equation} 
where $ y_k^i =[y_{k,1}^i,y_{k,2}^i]^{T}$ and $\ymeas{k}{l}{i}$ indicates the $l^{th}$ dimension of $y_k^i$ for $l\in\{1,2\}$.
 Following the idea of  \cite{Fusion12_Baum}  of modeling the measurement spread as multiplicative noise  and using our parameterization \Eq{eqn:param}, we can form the measurement equation
 \begin{align}\label{eqn:multmeas}
        y_k^i  &=  m_k +  \hnoise{k}{1}{i} \cdot  \xl{k}{1} \cdot \vect{\cos{\alpha_k}\\ \sin{\alpha_k}} + \hnoise{k}{2}{i} \cdot \xl{k}{2} \cdot \vect{-\sin{\alpha_k}\\ \cos{\alpha_k}}      + v_k^i \\
        & =: h(x_k,v_k^i, h^i_k) \notag 
       \end{align}
   with 
       \begin{itemize}
        \item zero-mean multiplicative (Gaussian) noise $h^i_k = \tvect{ \hnoise{k}{1}{i}, \hnoise{k}{2}{i} } \in \IR^2$   with covariance $\diag(c_1,c_2)$, where $c_1$ and $c_2$ are constant factors that specify the spread of the measurements on the object, and
        \item additive  Gaussian measurement noise $v_k^i$ with covariance $\mat{Q}_k^i$.
        \end{itemize}
        
Intuitively, $\hnoise{k}{1}{i}$ and $\hnoise{k}{2}{i}$ in \Eq{eqn:multmeas} randomly scale the semi-axis of the ellipse.

A   noise-free measurement ($v_k^i=\mat{0}$)  refers to its ``measurement source'', see \Fig{fig:ellipse_par}.  As the measurement source is supposed to lie on the ellipse, physical meaningful values of $\hnoise{k}{1}{i}$ and $\hnoise{k}{2}{i}$ should lie in $[-1 ,1 ]$, e.g., a uniform  distribution on $[-1, 1]$ would be reasonable.

 \subsection{Process Model}
 For the sake of simplicity, we focus on linear process models 
 \begin{equation}
 x_{k+1} = \mat{A}_k x_k +w_k  \enspace,
 \end{equation}
where 
\begin{itemize}
 \item $\mat{A}_k$ is the system matrix, and 
 \item $w_k$ is zero-mean white Gaussian process noise with covariance matrix $\mat{P}_k$.
\end{itemize}

\section{Relationship to the Random Matrix Measurement Model}
The random matrix approach introduced in \cite{Feldmann2008} employs the likelihood function
\begin{equation}\label{eqn:likelihood}
 p(\meas{k}{i}| m_k, \mat{X} ) \sim  \Gauss{\meas{k}{i}}{m_k}{c \mat{X}_k+ \mat{Q}^i_k} \enspace,
\end{equation}
where
\begin{itemize}
\item $\mat{X}_k$ is a symmetric positive definite matrix that specifies the elliptic extend,
\item $c\in \IR$ is a constant scaling factor, e.g., to match uniform measurement spread,
\item $\mat{Q}_k^i$ is the measurement noise covariance.
\end{itemize}
 Actually, the corresponding likelihood function of measurement equation \Eq{eqn:multmeas}  coincides with \Eq{eqn:likelihood} if $h_k^i$ is Gaussian distributed. Only the parameterization of the ellipse differs.
In order to show that, we first note that any covariance matrix $\mat{X}_k$ can be written as 
 \begin{equation}
  \mat{X}_k = \mat{R}_k \mat{D}_k \mat{R}_k^T\enspace,
 \end{equation}
with 
\begin{align}
\mat{R}_k &= \vect{\cos{\alpha_k} &  -\sin{\alpha_k} \\  \sin{\alpha_k} &  \cos{\alpha_k}  }\enspace,\\
 \mat{D}_k &= \vect{(\xl{k}{1})^2 &0  \\ 0 & (\xl{k}{2})^2 } \enspace .
\end{align} 
In this manner, \Eq{eqn:multmeas} can be written as
 \begin{align}\label{eqn:multmeas2}
        y_k^i  &=  m_k +  \mat{R}_k \sqrt{\mat{D}_k}\cdot h_k^i   + v_k^i   \enspace .
       \end{align}

\section{Second-Order Extended Kalman Filter}\label{sec:SOEKF}
In this section, we derive a second-order Kalman filter (SOEKF) for recursively estimating the kinematic and shape parameters of an extended object based on the models introduced in the previous section.
As we have to deal with multiple measurements per time step, we will process the measurements sequentially. For this purpose, let 
$\hat{x}^{i}_k$ and $\mat{C}^{i}_k$ denote the  mean and covariance of the estimate having incorporated all measurements up to the  $i$-th measurement  of time $k$.
According to this, notation $\hat{x}^{0}_k$ and $\mat{C}^{0}_k$ represent the prediction for time $k$, having not yet incorporated a measurement from time $k$.

\subsection{Measurement Update}
 As shown in our previous work \cite{Fusion12_Baum} for (axis-aligned) ellipses, the optimal \emph{linear} estimator is not feasible for the multiplicative noise model \Eq{eqn:multmeas} as there are not ``enough''  correlations between the measurement and state vector.
 Hence, we create a \emph{quadratic} estimator by forming a pseudo-measurement from the original measurement and the $2$-fold Kronecker product    
\begin{equation}\label{eqn:meas_kronecker}
(y_k^i)^{[2]} = \vect{(\ymeas{k}{1}{i})^2\\(\ymeas{k}{2}{i})^2  \\ \ymeas{k}{1}{i}\cdot  \ymeas{k}{2}{i} } \enspace.
\end{equation} 
  Furthermore, we shift the (estimated) center $\hat{m}_k^{i-1}$ of the object  to the origin in order to avoid numerical problems due to the squared equation. It is important to note that all these reformulations do not change the original  likelihood function. However, when using the Kalman filter update equations, an improvement can be  achieved as the squared measurements are incorporated. This concept is widely-known and frequently used in literature, see for example \cite{Fusion12_Baum,Carravetta1997,DeSantis1995,Lan2015,Liu2013,Wuethrich2015}. 
All told, the final measurement equation becomes
   \begin{equation}\label{eqn:quadr}
            \underbrace{ \vect{y_k^i-\hat{m}_k^{i-1}\\(y_k^i-\hat{m}_k^{i-1})^{[2]}}}_{:={z}_k^i} =  \underbrace{\vect{h(x_k,v_k^i, h^i_k)-\hat{m}_k^{i-1} \\  (h(x_k,v_k^i, h^i_k)-\hat{m}_k^{i-1})^{[2]} }}_{:=g(x_k,v_k^i, h^i_k)} \enspace ,
        \end{equation}
where operator $(\cdot)^{[2]}$ is the 2-fold Kronecker product as we defined in \Eq{eqn:meas_kronecker}. Based on  \Eq{eqn:quadr} the Kalman filter update becomes
 \begin{eqnarray}\label{eqn:kalmanfilter}
\hat{x}^{i}_k   &= &   \hat{x}^{i-1}_k   +  \mat{M}^i_k (\mat{S}^i_k)^{-1}  (z^{i}_{k}-  \bar{z}^{i}_{k} )   \enspace, \\
  \mat{C}^{i}_{k} &=&   \mat{C}^{i-1}_k     -\mat{M}^i_k (\mat{S}^i_k)^{-1} (\mat{M}^i_k)^T  \enspace ,   
\end{eqnarray}
with
\begin{eqnarray}
 \mat{M}^i_k &= \Cov{z_k^i, x_k \;  | \; \mathcal{Z}_{k}^{i-1}}\enspace, \label{eqn:cov_zx} \\
\mat{S}^i_k &= \Cov{z_k^i, z_k^i \;  | \; \mathcal{Z}_{k}^{i-1}}\enspace, \label{eqn:cov_zz} \\
\bar{z}^{i}_{k}  &=  \Exp{z_k^i \;  | \; \mathcal{Z}_{k}^{i-1}} \enspace. \label{eqn:e_z}
\end{eqnarray}

 It turned out that a first-order Taylor series approximation of \Eq{eqn:quadr}  is not precise enough to capture all nonlinearities.
 Hence, we propose a second-order Taylor series approximation \cite{Roth2011}.  
 If we define the augmented state vector $\gamma_{k} = \tvect{x_k^T,(v_k^i)^T, (h^i_k)^T}$
with $\hat{\gamma}_{k}^i =  \tvect{(\hat{x}^i_k)^T,0, 0, 0, 0}$ and covariance $\mat{\Gamma}_k^{i-1} = \diag(\mat{C}^{i-1}_k,\mat{Q}_k^i,c_1,c_2)$,
we obtain  \cite{Roth2011}
\begin{align}
 \Exp{z_{k,l}^i \;  | \; \mathcal{Z}_{k}^{i-1}} &=  g_l(\hat{\gamma}_{k}^{i-1})+ \frac{1}{2} \Tr(\mat{H}_{k,l}^{i-1} \mat{\Gamma}_k^{i-1}) \enspace,\\
 \Cov{z_{k,l}^i, z_{k,r}^i \;  | \; \mathcal{Z}_{k}^{i-1}}  &=  \mat{J}_{k,l}^{i-1} \mat{\Gamma}_k^{i-1} (\mat{J}_{k,r}^{i-1})^T \notag\\ &\phantom{=}  + \frac{1}{2} \Tr(\mat{H}_{k,l}^{i-1} \mat{\Gamma}_k^{i-1} \mat{H}_{k,r}^{i-1} \mat{\Gamma}_k^{i-1})\enspace,\\
 \Cov{z_k^i, \gamma_k \;  | \; \mathcal{Z}_{k}^{i-1}}  &= \mat{\Gamma}_k^{i-1} (\mat{J}_k^{i-1})^T\enspace,
\end{align}
where 
\begin{itemize}
\item $\mat{J}_{k}^i$ is the Jacobian matrix of  $g$ evaluated at $\hat{\gamma}_{k}^i$,  $\mat{J}_{k,l}^i$ denotes the $l$-th row of $\mat{J}_{k}^i$, and
\item $\mat{H}_{k,l}^i$ is the Hessian matrix of the $l$-th component function of $g$ evaluated at $\hat{\gamma}_{k}^i$. 
\end{itemize}
The Jacobian and Hessians are given in the Appendix.
We note that an essential modification of the Jacobian and Hessians is necessary:
As the means of $h^i_{k,1}$ and $h^i_{k,2}$ are $0$,  significant parts of  Jacobian and Hessians at $\hat{\gamma}_{k}^i$ are zero as well.
Hence, we substitute $(h^i_{k,1})^2$ and $(h^i_{k,2})^2$ in the Jacobian and Hessians   by $\Exp{(h^i_{k,1})^2}=c_1$ and $\Exp{(h^i_{k,2})^2}=c_2$.
 Without this modifications, the shape parameters do not change in  a measurement update. 

\subsection{Time Update}
As the process model is linear and  the time update can be performed  with the standard Kalman filtering equations.

\section{Evaluation}
In this section, we first briefly discuss the current approaches used for extend object tracking and suggest a new metric based on a Wasserstein/Optimal Sub-Pattern Assignment (OSPA) distance \cite{Schuhmacher2008} construction. Then, we evaluate our method for tracking elliptical and rectangular objects in static and dynamic scenarios using suggested metric. In both simulations, we compare our proposed SOEKF estimator with a Monte Carlo approximation.

\begin{figure}\centering
\includegraphics[scale=0.8]{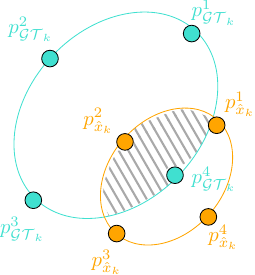}
\caption{An illustration for the points we selected for evaluation. The red ellipse is estimation $\hat{x}_k$ and the green ellipse indicates ground truth $\mathcal{GT}_k$ in time step $k$.}
\label{fig:ellipse_points}
\end{figure}

For extended object tracking,  the  state vector normally includes   kinematic and shape parameters \cite{Koch2008}. 
As these quantities are not at the same order of magnitude and different shape parameters can specify the same shape, the overall Root Mean Squared Error (RMSE) of the estimated state would be misleading. This problem can be by-passed by decoupling state properties and calculate their RMSEs separately \cite{Fusion12_Baum,Feldmann2010}. Decoupled RMSE gives a more detailed insight for the performance in a certain aspect. 
To combine object shape, size, and orientation, a similarity measure called Intersection-over-Union (IoU) \cite{Granstroem2011}, which is also known as Jaccard index,  is widely used in the evaluation of many computer vision tasks, such as image segmentation  \cite{Everingham2015,Csurka2013}, object detection \cite{Nascimento2006,Zitnick2008} and tracking \cite{Breuers2016,Milan2013}. 
Given two shapes, IoU is the intersected area divided by their union area. 
Using IoU to evaluate extended object tracking methods still has two major drawbacks. 
Firstly, IoU is extremely difficult to calculate for  non-axis aligned objects as the intersection and union areas are normally irregular shapes (see Fig.~\ref{fig:ellipse_points}). 
In computer vision, IoU  is typically  calculated either for regular axis-aligned objects or approximated using the number of intersected pixels divided by the number of pixels on the union area.  Secondly, even if we could approximate the area of intersection and union by sampling \cite{Granstroem2011}, IoU score could not distinguish two estimates when neither of them intersect with the ground truth. 

Based on the discussion above, we suggest a miss-distance for extended object tracking evaluation based on the  Wasserstein/OSPA distance. 
Rectangular and elliptical objects have two axes of symmetry that intersect the ellipse contour at four points. 
These four points capture differences in position, shape, size, and orientation, hence, uniquely determine a rectangle or ellipse.
We select aforementioned four points from  the ground truth ($\mathcal{GT}_k$) and the estimate $\hat{x}_k$ in time step $k$. 
This gives two sets of four points (see Fig.~\ref{fig:ellipse_points}), $\mathbb{S}_{\mathcal{GT}_{k}}=\{p_{\mathcal{GT}_{k}}^1,p_{\mathcal{GT}_{k}}^2,p_{\mathcal{GT}_{k}}^3,p_{\mathcal{GT}_{k}}^4\}$ and $\mathbb{S}_{\hat{x}_{k}}=\{p_{\hat{x}_{k}}^1, p_{\hat{x}_{k}}^2, p_{\hat{x}_{k}}^3, p_{\hat{x}_{k}}^4\}$, whose 
distance can be calculated with the Wasserstein/OSPA according to
\begin{equation}
\mathbf{d}_{EOT}(\hat{x}_k, \mathcal{GT}_k) = \min_{\pi \in \Pi}\sqrt{\frac{1}{4}\sum_{i=1}^{4}\parallel p_{\mathcal{GT}_{k}}^i - p_{\hat{x}_{k}}^{\pi (i)}\parallel ^2} \enspace ,
\label{eqn:error_function}
\end{equation}
where $\Pi$ is the set of all permutations of $\{1,2,3,4\}$. 
For perfectly aligned estimate and ground truth, $\mathbf{d}_{EOT}$ is 0, i.e., no estimation error. It it obvious that $\mathbf{d}_{EOT}$ satisfies the identity, symmetry, and triangle inequality that a metric requires.
Besides, $\mathbf{d}_{EOT}$ could also compare two estimates even when neither of them intersect with the ground truth.

\subsection{Stationary Ellipse}
As we derived a SOEKF for a closed-form  measurement update in Section \ref{sec:SOEKF}, we would like to evaluate its performance  compared to Monte Carlo sampling for the moment matching in \eqref{eqn:cov_zx}, \eqref{eqn:cov_zz}, and \eqref{eqn:e_z} using 10000 samples.
In order to focus on the measurement update, we first consider a stationary object.

\begin{figure*}[!t]
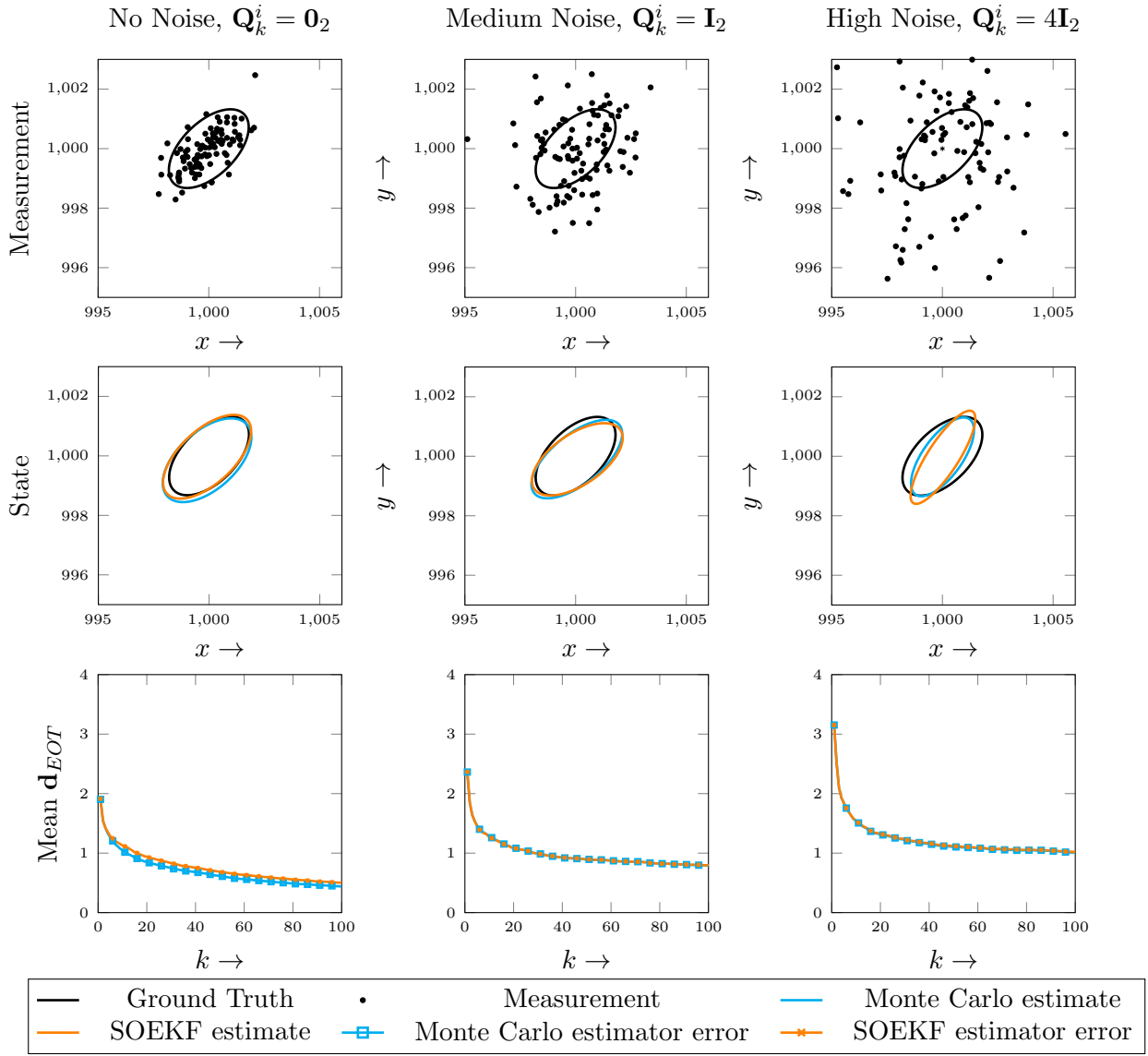
\centering
	\begin{tikzpicture}
	\tikzset{mark size=1,line width = 1pt}
		\begin{groupplot}[group style = {group size = 3 by 3, horizontal sep = 50pt}, width = 0.3\textwidth, height = 5.0cm]
			\nextgroupplot[ title = {No Noise, $\mat{Q}^{i}_k=\mat{0}_2$},ylabel={Measurement},xlabel={$x\rightarrow$},
							ymin=995,ymax=1003,xmin=995,xmax=1006]
				\input{static_meas_meas_noise_0.tex}
				\label{fig:static_meas_meas_noise_0}
			\nextgroupplot[title = {Medium Noise, $\mat{Q}^{i}_k=\mat{I}_2$},
							xlabel={$x\rightarrow$},ylabel={$y\rightarrow$},ymin=995,ymax=1003,xmin=995,xmax=1006]
				\input{static_meas_meas_noise_1.tex}
				\label{fig:static_meas_meas_noise_1}
			\nextgroupplot[title = {High Noise, $\mat{Q}^{i}_k=4\mat{I}_2$},xlabel={$x\rightarrow$},ylabel={$y\rightarrow$},
							ymin=995,ymax=1003,xmin=995,xmax=1006]
				\input{static_meas_meas_noise_2.tex}
				\label{fig:static_meas_meas_noise_2}
				
			\nextgroupplot[ylabel={State},xlabel={$x\rightarrow$},
							legend style = { column sep = 10pt, legend columns = 3, legend to name = grouplegend,},
							ymin=995,ymax=1003,xmin=995,xmax=1006]
				\input{static_state_meas_noise_0.tex}
				\label{fig:static_state_meas_noise_0}
			\nextgroupplot[xlabel={$x\rightarrow$},ylabel={$y\rightarrow$},
							ymin=995,ymax=1003,xmin=995,xmax=1006]
				\input{static_state_meas_noise_1.tex}
				\label{fig:static_state_meas_noise_1}
			\nextgroupplot[xlabel={$x\rightarrow$},ylabel={$y\rightarrow$},
							ymin=995,ymax=1003,xmin=995,xmax=1006]
				\input{static_state_meas_noise_2.tex}
				\label{fig:static_state_meas_noise_2}

			\nextgroupplot[ylabel={Mean $\mathbf{d}_{EOT}$},xlabel={$k\rightarrow$},ymin=0,ymax=4,xmin=0,xmax=100]
				\input{static_rmse_meas_noise_0.tex}
				\label{fig:static_rmse_meas_noise_0}
			\nextgroupplot[xlabel={$k\rightarrow$},,ymin=0,ymax=4,xmin=0,xmax=100]
				\input{static_rmse_meas_noise_1.tex}
				\label{fig:static_rmse_meas_noise_1}
			\nextgroupplot[xlabel={$k\rightarrow$},,ymin=0,ymax=4,xmin=0,xmax=100]
				\input{static_rmse_meas_noise_2.tex}
				\label{fig:static_rmse_meas_noise_2}
		\end{groupplot}
		\node at ($(group c2r3) + (0,-3.2cm)$) {\ref{grouplegend}}; 
	\end{tikzpicture}
	\caption{Simulation results for a static ellipse. The first row shows the ground truth and measurements. The second row gives example estimates after $100$ measurement update. The last row shows mean $\mathbf{d}_{EOT}$ for $100$ runs.}
		\label{fig:static_ellipse_result}
\end{figure*}

The ground truth ellipse lies at $\tilde{m}_k = [1000,1000]$, $30\,^{\circ}$ counter-clockwise rotated from the $x$-axis, i.e., $\tilde{\alpha}_k = \frac{\pi}{6}$, and the length is $\tilde{l}_k = [2,1]^{T}$, for all $k$.  
As described in our previous work \cite{Fusion12_Baum}, $h_k^i$ lies on the interval of $[-1,1]$. To ensure that most measurement sources lie on the object extent, the multiplicative noise $h_k^i$ follows Gaussian distribution $\mathcal{N}(h_k^i-\mat{0}_2,\frac{1}{4}\mat{I}_2)$ \cite{Feldmann2010}.

 We test our approach under three scenarios: no measurement noise ($\mat{Q}^{i}_k=\mat{0}_2$), medium measurement noise ($\mat{Q}^{i}_k=\mat{I}_2$), and high measurement noise ($\mat{Q}^{i}_k=4\mat{I}_2$). 
The prior is given by a Gaussian distribution with $9\mat{I}$ as covariance of center, $\frac{1}{9}$ as variance of orientation, and $\mat{I}_2$ as covariance matrix of lengths.

The measurements, example estimates and mean error for the  described simulations are shown in Fig.~\ref{fig:static_ellipse_result}.
We can see that the proposed SOEKF estimator is slightly worse than Monte Carlo sampling when there is no measurement noise. However, it coincides with Monte Carlo under medium and high measurement noise.  
The simulations show that the SOEKF gives pretty good approximations for the moments  in \Eq{eqn:kalmanfilter}, even though the degree of the measurement equation is much higher than two.

\subsection{Rectangle with NCV}
\begin{figure*}[!t]
\centering
\tikzset{mark size=0.5}
\input{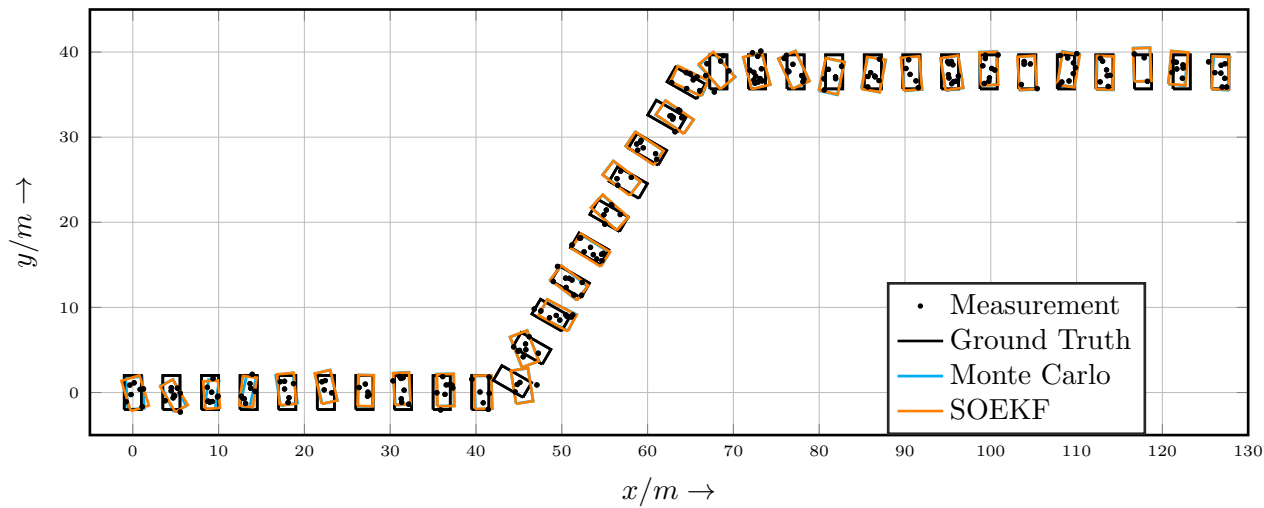}
\caption{Example tracking result for a rectangle. Ground truth and estimates are plotted for every third time step.}
\label{fig:dyn_rectangle}
\end{figure*}

In the following, we evaluate our method for tracking a rectangular object that follows a Nearly Constant Velocity (NCV) model. 

The object initially lies at the origin, the width of the object is $2m$ and the length is $4m$, i.e., $\tilde{l}_k = [1,2]^{T}$ for all $k$. 
The object orientation is aligned with its velocity. 
For the first $30$ and last $40$ time steps, the object moves along $x$ axes and the speed  is $1.5m$ per time step.
 In between, its velocity is $[1.5,1.5]^{T}$. 
The number of measurements in each time step is drawn from a Poisson distribution with mean of $7$.
The measurement sources are uniformly distributed on the extent of the object, which results in  $h_k^i \sim \mathcal{U}(h_k^i-\mat{0}_2,\frac{1}{3}\mat{I}_2)$.
The measurement noise is zero-mean Gaussian distributed with covariance of $\frac{1}{3}\mat{I}_2$. 
 The initial guess are  $m_0=[0.6,0.6]^T$ with covariance of $\frac{1}{2}\mat{I}_2$, $\alpha_0 =\frac{\pi}{3}$ with variance of $0.76$, $l_0=[1.5,2.5]$ with covariance of $\frac{1}{5}\mat{I}_2$, and velocity $[1,0]^T$ with covariance matrix $\mat{I}_2$.  
The  ground truth and an example estimation result  is depicted in Fig.~\ref{fig:dyn_rectangle}  for every third time step.
 Consistent with the results from the static case, SOEKF estimations overlap with Monte Carlo estimations after sufficient number of measurements. 
All told, the simulations demonstrate that the second-order approximation   is very accurate  even in the case of high noise.

\section{Conclusion and Future Work}
Simultaneous tracking and shape estimation based on independent  noisy point measurements is  a challenging nonlinear estimation problem -- even for basic shapes such as ellipses.
This work started from the idea to develop  a standard nonlinear Gaussian estimator for estimating an elliptic shape approximation.

It turned out that three steps are required: (i) A measurement function with multiplicative noise must be formulated. (ii) The measurement space needs to be augmented; otherwise there are not enough  correlations between the measurements and shape parameters. (iii) A first-order Taylor series expansion of the quadratic measurement equation is not sufficient. However, a second-order  Taylor series expansion  (SOEKF)  pretty much matches  the
exact moments (but only if we substitute the mean of the squared multiplicative noise).

The final equations of the SOEKF are still tractable and rather compact. However, we believe that significant simplifications are possible, e.g., if the kinematic and shape parameters are assumed to be independent.

\begin{appendix}
\section{Jacobian and Hessian matrices}
\noindent

This Appendix gives the Jacobian and Hessian matrix for our SOEKF estimator in \Fig{fig:jac}. For compactness, we 
\begin{itemize}
\item suppress the time index $k$, measurement index $i$, and
\item omit the kinematic parameters $r_k$ as they do not appear in the measurement equation
\item do not differentiate the spread of multiplicative error, i.e., $c_1=c_2=c$.
\end{itemize}

\begin{figure*}[!t]\centering
\frame{
\begin{tabular}{ccl}
 $\mat{J}_{\cdot}^{\cdot}$&=&
 \scalebox{0.7}{ $
\left(\begin{array}{ccccccccc} 
1 & 0 & 0 & 0 & 0 & l_1\, \cos{\alpha} & - l_2\, \sin{\alpha} & 1 & 0\\
 0 & 1 & 0 & 0 & 0 & l_1\, \sin{\alpha} & l_2\, \cos{\alpha} & 0 & 1\\
  0 & 0 & - c\, \sin{2\alpha}\, \left({l_1}^2 - {l_2}^2\right) & 2\, l_1\, c\, \cos^{2}\alpha & 2\, l_2\, c\, \sin^{2}\alpha & 0 & 0 & 0 & 0\\ 0 & 0 & c\, \sin{2\alpha}\, \left({l_1}^2 - {l_2}^2\right) & 2\, l_1\, c\, \sin^{2}\alpha & 2\, l_2\, c\, \cos^{2}\alpha & 0 & 0 & 0 & 0\\ 0 & 0 & c\, \cos{2\alpha}\, \left({l_1}^2 - {l_2}^2\right) & l_1\, c\, \sin{2\alpha} & - l_2\, c\, \sin{2\alpha} & 0 & 0 & 0 & 0 \end{array}\right)
  $}\\
 $ \mat{H}_{\cdot,1}^{\cdot}$&=&
  \scalebox{0.7}{ $
\left(\begin{array}{ccccccccc} 0 & 0 & 0 & 0 & 0 & 0 & 0 & 0 & 0\\ 0 & 0 & 0 & 0 & 0 & 0 & 0 & 0 & 0\\ 0 & 0 & 0 & 0 & 0 & - l_1\, \sin{\alpha} & - l_2\, \cos{\alpha} & 0 & 0\\ 0 & 0 & 0 & 0 & 0 & \cos{\alpha} & 0 & 0 & 0\\ 0 & 0 & 0 & 0 & 0 & 0 & - \sin{\alpha} & 0 & 0\\ 0 & 0 & - l_1\, \sin{\alpha} & \cos{\alpha} & 0 & 0 & 0 & 0 & 0\\ 0 & 0 & - l_2\, \cos{\alpha} & 0 & - \sin{\alpha} & 0 & 0 & 0 & 0\\ 0 & 0 & 0 & 0 & 0 & 0 & 0 & 0 & 0\\ 0 & 0 & 0 & 0 & 0 & 0 & 0 & 0 & 0 \end{array}\right)
$}\\
  $\mat{H}_{\cdot,2}^{\cdot}$&=&
 \scalebox{0.7}{ $\left(\begin{array}{ccccccccc} 0 & 0 & 0 & 0 & 0 & 0 & 0 & 0 & 0\\ 0 & 0 & 0 & 0 & 0 & 0 & 0 & 0 & 0\\ 0 & 0 & 0 & 0 & 0 & l_1\, \cos{\alpha} & - l_2\, \sin{\alpha} & 0 & 0\\ 0 & 0 & 0 & 0 & 0 & \sin{\alpha} & 0 & 0 & 0\\ 0 & 0 & 0 & 0 & 0 & 0 & \cos{\alpha} & 0 & 0\\ 0 & 0 & l_1\, \cos{\alpha} & \sin{\alpha} & 0 & 0 & 0 & 0 & 0\\ 0 & 0 & - l_2\, \sin{\alpha} & 0 & \cos{\alpha} & 0 & 0 & 0 & 0\\ 0 & 0 & 0 & 0 & 0 & 0 & 0 & 0 & 0\\ 0 & 0 & 0 & 0 & 0 & 0 & 0 & 0 & 0 \end{array}\right)$}\\
  $\mat{H}_{\cdot,3}^{\cdot}$&=&
 \scalebox{0.7}{$\left(\begin{array}{ccccccccc} 2 & 0 & 0 & 0 & 0 & 2\, l_1\, \cos{\alpha} & - 2\, l_2\, \sin{\alpha} & 2 & 0\\ 0 & 0 & 0 & 0 & 0 & 0 & 0 & 0 & 0\\ 0 & 0 & - 2\, c\, \cos{2\alpha}\, \left({l_1}^2 - {l_2}^2\right) & - 2\, l_1\, c\, \sin{2\alpha} & 2\, l_2\, c\, \sin{2\alpha} & 0 & 0 & 0 & 0\\ 0 & 0 & - 2\, l_1\, c\, \sin{2\alpha} & 2\, c\, \cos^{2}\alpha & 0 & 0 & 0 & 0 & 0\\ 0 & 0 & 2\, l_2\, c\, \sin{2\alpha} & 0 & 2\, c\, \sin^{2}\alpha & 0 & 0 & 0 & 0\\ 2\, l_1\, \cos{\alpha} & 0 & 0 & 0 & 0 & 2\, {l_1}^2\, \cos^{2}\alpha & - l_1\, l_2\, \sin{2\alpha} & 2\, l_1\, \cos{\alpha} & 0\\ - 2\, l_2\, \sin{\alpha} & 0 & 0 & 0 & 0 & - l_1\, l_2\, \sin{2\alpha} & 2\, {l_2}^2\, \sin^{2}\alpha & - 2\, l_2\, \sin{\alpha} & 0\\ 2 & 0 & 0 & 0 & 0 & 2\, l_1\, \cos{\alpha} & - 2\, l_2\, \sin{\alpha} & 2 & 0\\ 0 & 0 & 0 & 0 & 0 & 0 & 0 & 0 & 0 \end{array}\right)$}\\
  $\mat{H}_{\cdot,4}^{\cdot}$&=&
 \scalebox{0.7}{ $\left(\begin{array}{ccccccccc} 0 & 0 & 0 & 0 & 0 & 0 & 0 & 0 & 0\\ 0 & 2 & 0 & 0 & 0 & 2\, l_1\, \sin{\alpha} & 2\, l_2\, \cos{\alpha} & 0 & 2\\ 0 & 0 & 2\, c\, \cos{2\alpha}\, \left({l_1}^2 - {l_2}^2\right) & 2\, l_1\, c\, \sin{2\alpha} & - 2\, l_2\, c\, \sin{2\alpha} & 0 & 0 & 0 & 0\\ 0 & 0 & 2\, l_1\, c\, \sin{2\alpha} & 2\, c\, \sin^{2}\alpha & 0 & 0 & 0 & 0 & 0\\ 0 & 0 & - 2\, l_2\, c\, \sin{2\alpha} & 0 & 2\, c\, \cos^{2}\alpha & 0 & 0 & 0 & 0\\ 0 & 2\, l_1\, \sin{\alpha} & 0 & 0 & 0 & 2\, {l_1}^2\, \sin^{2}\alpha & l_1\, l_2\, \sin{2\alpha} & 0 & 2\, l_1\, \sin{\alpha}\\ 0 & 2\, l_2\, \cos{\alpha} & 0 & 0 & 0 & l_1\, l_2\, \sin{2\alpha} & 2\, {l_2}^2\, \cos^{2}\alpha & 0 & 2\, l_2\, \cos{\alpha}\\ 0 & 0 & 0 & 0 & 0 & 0 & 0 & 0 & 0\\ 0 & 2 & 0 & 0 & 0 & 2\, l_1\, \sin{\alpha} & 2\, l_2\, \cos{\alpha} & 0 & 2 \end{array}\right)$}\\
  $\mat{H}_{\cdot,5}^{\cdot}$&=&
 \scalebox{0.7}{ $\left(\begin{array}{ccccccccc} 0 & 1 & 0 & 0 & 0 & l_1\, \sin{\alpha} & l_2\, \cos{\alpha} & 0 & 1\\ 1 & 0 & 0 & 0 & 0 & l_1\, \cos{\alpha} & - l_2\, \sin{\alpha} & 1 & 0\\ 0 & 0 & - 2\, c\, \sin{2\alpha}\, \left({l_1}^2 - {l_2}^2\right) & 2\, l_1\, c\, \cos{2\alpha} & - 2\, l_2\, c\, \cos{2\alpha} & 0 & 0 & 0 & 0\\ 0 & 0 & 2\, l_1\, c\, \cos{2\alpha} & c\, \sin{2\alpha} & 0 & 0 & 0 & 0 & 0\\ 0 & 0 & - 2\, l_2\, c\, \cos{2\alpha} & 0 & - c\, \sin{2\alpha} & 0 & 0 & 0 & 0\\ l_1\, \sin{\alpha} & l_1\, \cos{\alpha} & 0 & 0 & 0 & {l_1}^2\, \sin{2\alpha} & l_1\, l_2\, \cos{2\alpha} & l_1\, \sin{\alpha} & l_1\, \cos{\alpha}\\ l_2\, \cos{\alpha} & - l_2\, \sin{\alpha} & 0 & 0 & 0 & l_1\, l_2\, \cos{2\alpha} & - {l_2}^2\, \sin{2\alpha} & l_2\, \cos{\alpha} & - l_2\, \sin{\alpha}\\ 0 & 1 & 0 & 0 & 0 & l_1\, \sin{\alpha} & l_2\, \cos{\alpha} & 0 & 1\\ 1 & 0 & 0 & 0 & 0 & l_1\, \cos{\alpha} & - l_2\, \sin{\alpha} & 1 & 0 \end{array}\right)$}
\end{tabular}
}
\caption{Jacobian matrix and Hessian matrices for the SOEKF. Note that the Hessian matrices are symmetric. \label{fig:jac}}
\end{figure*}

\end{appendix}

\bibliographystyle{IEEEtranS}
\bibliography{../../../BibTex/Literature,../../../BibTex/publicationsFusion.bib}

\end{document}